\documentclass{article}
\usepackage{emulateapj,timesfonts}

\long\def\***#1{{\scshape ***#1***}}
\def\arraystretch{1.25}

\makeatletter
\newenvironment{inlinetable}{%
\def\@captype{table}%
\noindent\begin{minipage}{0.999\linewidth}\begin{center}\footnotesize}
{\end{center}\end{minipage}\smallskip}

\newenvironment{inlinefigure}{%
\def\@captype{figure}%
\noindent\begin{minipage}{0.999\linewidth}\begin{center}}
{\end{center}\end{minipage}\smallskip}
\makeatother

\def\Lx{L_{\rm x}}
\def\ML{M\!/\!L}
\def\m{\ensuremath{^{\rm m}}}

\begin{document}
\lefthead{VIKHLININ ET AL.}
\righthead{OVER-LUMINOUS ELLIPTICAL GALAXIES}
\submitted{Submitted to ApJ Letters, February 26, 1999}

\title{X-ray Over-Luminous Elliptical Galaxies: \\ A New Class of Mass
  Concentrations in the Universe?}  \author{A.~Vikhlinin\altaffilmark{1},
  B.~R.~McNamara\altaffilmark{1}, A.~Hornstrup\altaffilmark{2},
  H.~Quintana\altaffilmark{3}, W.~Forman\altaffilmark{1},
  C.~Jones\altaffilmark{1}, M.~Way\altaffilmark{4}} 

\altaffiltext{1}{Harvard-Smithsonian Center for Astrophysics, 60 Garden St.,
  Cambridge, MA}

\altaffiltext{2}{Danish Space Research Institute, Copenhagen, Denmark}

\altaffiltext{3}{Dpto.\ de Astronomia y Astrofisica, Pontificia Universidad
  Catolica, Santiago, Chile}

\altaffiltext{4}{Department of Astronomical Sciences, Princeton University}

\begin{abstract}
  We detect four isolated, X-ray over-luminous
  ($\Lx>2\times10^{43}\,[h/0.5]^{-2}$\,erg~s$^{-1}$) elliptical galaxies
  (OLEGs) in our 160 square degree \emph{ROSAT} PSPC survey. The extent of
  their X-ray emission, total X-ray luminosity, total mass, and mass of the
  hot gas in these systems correspond to poor clusters, and the optical
  luminosity of the central galaxies ($M_R<-22.5+5\lg h$) is comparable to
  that of cluster cDs.  However, there are no detectable fainter galaxy
  concentrations around the central elliptical. The mass-to-light ratio
  within the radius of detectable X-ray emission is in the range
  $250-450M_\odot/L_\odot$, which is 2--3 times higher than typically found
  in clusters or groups. These objects can be the result of galaxy merging
  within a group. However, their high $M/L$ values are difficult to explain
  in this scenario. OLEGs must have been undisturbed for a very long time,
  which makes them the ultimate examples of systmes in hydrostatic
  equilibrium. The number density of OLEGs is
  $n=2.4^{+3.1}_{-1.2}\times10^{-7}\, (h/0.5)^{-3}$~Mpc$^{-3}$ at the 90\%
  confidence.  They comprise 20\% of all clusters and groups of comparable
  X-ray luminosity, and nearly all galaxies brighter than $M_R=-22.5$. The
  estimated contirubution of OLEGs to the total mass density in the Universe
  is close to that of $T>7\,$keV clusters.
%

\end{abstract}


\section{Introduction}

Large concentrations of matter in the Universe are found using optical
galaxies as tracers of mass. Systems with a wide range of mass and size were
discovered by this technique, from pairs and triplets of galaxies to
filaments extending for hunderds of Mpc. Do optical galaxy surveys detect
all large-scale mass concentrations, or do there exist populations of
``dark'' massive objects?

One approach for detection of dark systems is through gravitational lensing.
Week lensing observations can detect very large scale mass structures (e.g.,
Schneider et al.\ 1998), but this approach remains technically challenging.
Compact, massive systems near the line of sight of a distant object can be
detected by strong lensing. At least one such object was found: a massive
X-ray cluster at $z=1$ which is very poor optically for its mass (Hattori et
al.\ 1997, but see also Benitez et al.\ 1998).

X-ray surveys are another avenue for finding compact, massive, optically
dark objects, because such objects are likely to contain gravitationally
heated X-ray emitting gas. Tucker, Tananbaum \& Remillard (1995) did not
find any completely dark X-ray clusters in the \emph{Einstein} data.
However, X-ray surveys with \emph{ROSAT} did find a new class of objects
that could not be discovered optically --- bright isolated elliptical
galaxies surrounded by dark matter and hot gas halos typical of a group or
poor cluster. The first such object was a ``fossil group'' found by Ponman
et al.\ (1994).  The object appears optically as a giant isolated elliptical
galaxy. Its X-ray halo extends for at least 500~kpc; such an extent implies
the group-like total mass.  B\"ohringer et al.\ (1998) mention the existence
of similar objects in the \emph{ROSAT} All-Sky survey.  X-ray observation of
an optically selected isolated elliptical revealed the existence of a
group-like X-ray halo (Mulchaey \& Zabludoff 1998).

Our large-area \emph{ROSAT} survey of extended X-ray sources (Vikhlinin et
al.\ 1998, Paper~I hereafter) is ideal for the search for optically dark
clusters and groups at low redshift. In this \emph{Letter}, we report a
detection of four objects similar to the Ponman et al.\ fossil group. We
call them X-ray Over-Luminous Elliptical Galaxies (OLEGs).  We show that
these objects have gas and dark matter halos extending up to radii of 1~Mpc;
they are more numerous and massive than was previously appreciated, and
represent an important class of mass concentrations in the Universe.

We use $h=H_0/100\mbox{~km~s$^{-1}$~Mpc$^{-1}$}=0.5$, except where the
$h$-dependence is explicitly given, and $q_0=0.5$. All uncertainties are
reported as 68\% confidence intervals.

\begin{table*}
\caption{X-ray and optical properties of OLEGs}\label{tab:sample}
\def\x{\times}

\footnotesize
{\centering 
\begin{tabular}{p{2.5cm}cccccccccc}
\hline
\hline
& & $\Lx$ & $r_c$\textsuperscript{a}& $R_{\rm X}$\textsuperscript{b}& $T^{\rm(est)}$ & $t^{\rm cool}_0$ 
& $M_{\rm tot}^{\rm(est)}$ & $M_{\rm gas}^{\rm(est)}$ &$L_{\rm opt}$ & $\ML$ 
\\
\multicolumn{1}{c}{\raisebox{1.6ex}[0pt][0pt]{Object}}
& \raisebox{1.6ex}[0pt][0pt]{$z$}    &
 (erg~s$^{-1}$) & (kpc) & (kpc) & (keV) & (yr) &($M_\odot$) & ($M_\odot$) &
($L_\odot$) & ($M_\odot/L_\odot$) \\
\hline
$1159+5531$\dotfill&0.081&$2.2\x10^{43}$&50 &1000&2.2&$2.0\x10^{9\phantom{0}}$& $1.6\x10^{14}$&$9.2\x10^{12}$&$4.6\x10^{11}$ & 347 \\
$1340+4017$\dotfill&0.171&$2.5\x10^{43}$&71 &500 &2.3&$3.3\x10^{9\phantom{0}}$& $8.3\x10^{13}$&$4.8\x10^{12}$&$3.1\x10^{11}$ & 267 \\
$2114-6800$\dotfill&0.130&$2.0\x10^{43}$&61 &600 &2.1&$2.1\x10^{9\phantom{0}}$& $9.2\x10^{13}$&$5.1\x10^{12}$&$2.7\x10^{11}$ & 340 \\
$2247+0337$\dotfill&0.199&$4.1\x10^{43}$&194&850 &2.8&$1.2\x10^{10}$&$1.7\x10^{14}$&$1.6\x10^{13}$&$4.0\x10^{11}$ & 425 \\
\hline
\end{tabular}
\medskip

\begin{minipage}{0.87\linewidth}
\textsuperscript{a}--- X-ray core-radius. 
\textsuperscript{b}--- The radius of detectable X-ray emission.

{\sc Note} --- Masses are estimated within $R_{\rm X}$. Optical luminosities
are measured in the $R$ band within the same radius.
\end{minipage}
\par
}
\end{table*}

\section{Sample and Overall X-ray and Optical Properties}

The $R$ and $V$ band optical CCD images of serendipitously detected extended
X-ray sources from our \emph{ROSAT} PSPC catalog (Paper~I) were obtained on
the FLWO~1.2m and Danish~1.54m telescopes. We examined these images and
selected those objects that were obviously associated with a bright
elliptical galaxy, but had no corresponding concentration of fainter
galaxies. To add confidence to this selection, two further critera were
applied.  First, the central galaxy redshift was required to be $z<0.2$
because galaxy concentration is harder to detect in more distant objects.
Second, we required the 0.5--2~keV X-ray luminosity to be
$\Lx>2\times10^{43}\,$erg~s$^{-1}$. This luminosity corresponds to poor
Abell clusters which should be visible as galaxy concentrations in our CCD
images. X-ray fainter objects can be poor groups and therefore we could
misidentify them with single, isolated galaxies.  These selections resulted
in a sample of four objects (Table~\ref{tab:sample}).

Two of them, $1159+5531$ and $2114-6800$, were previously detected in the
\emph{Einstein} surveys (Stocke et al.\ 1991 and Griffiths et al. 1992,
respectively) and identified with normal elliptical galaxies. We also
rediscovered the Ponman et al.\ object, $1340+4017$. The remaining object,
$2247+0337$, first appears in our sample. We have measured the redshift of
$2247+0337$ and adopted the redshifts of others from the literature.  There
are several other plausible OLEG candidates, however they do not satisfy
either the redshift or the X-ray luminosity criteria.

In all four objects, the X-ray emission is detected with $2-3\sigma$
significance out to a large radius ${R_{\rm x}= 0.5-1\,}$Mpc
(Table~\ref{tab:sample}). The X-ray surface brightness distribution shows no
significant deviations from azimuthal symmetry.

Our optical images cover an area large enough to observe the entire X-ray
emitting region and to determine the local density of background galaxies.
We have measured the optical luminosity by co-adding light from the central
galaxy and all fainter galaxies within $R_{\rm x}$ and subtracting the
estimated background contribution. The resulting $R$-band luminosities (with
K- and extinction corrections applied) are in the range
$2.7-4.6\times10^{11}\,L_\odot$.  These values are uncertain by $\approx
25\%$ due to background fluctuations caused by the presence of relatively
bright foreground or background galaxies. In all objects, at least $70\%$ of
light (exluding the background) comes from the central galaxy; the
contribution from other galaxies does not exceed the level of background
fluctuations.

\begin{inlinefigure}
\bigskip
\centerline{\includegraphics[width=0.83\linewidth]{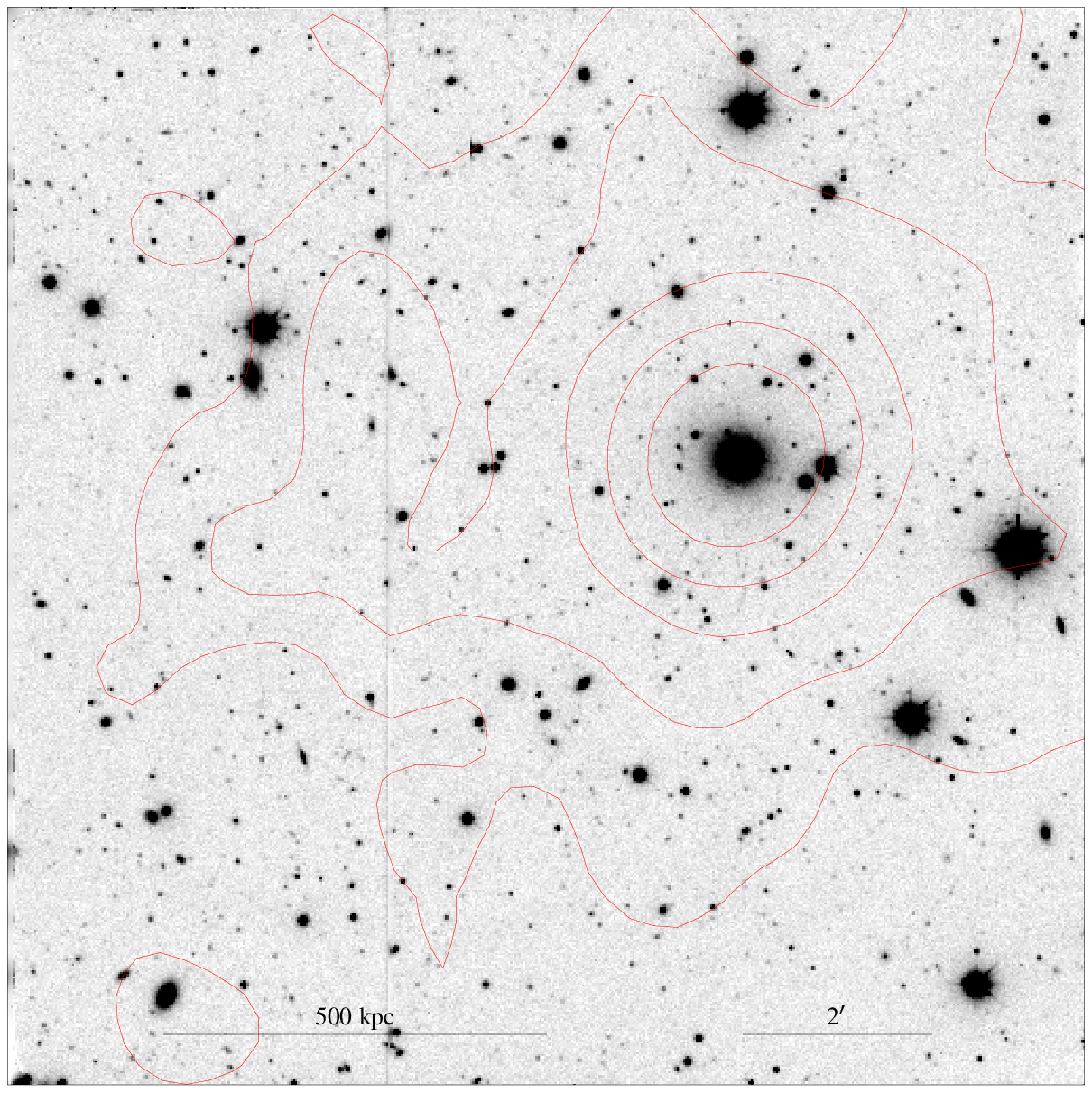}}
\bigskip
\caption{$R$-band CCD image of 1159+5531. The outer X-ray contour is
  approximately 800~kpc from the galaxy. Other bright objects are
  stars.}\label{fig:1159:opt}
\end{inlinefigure}

\section{Properties of 1159+5531}
\label{sec:1159}

Although this object was discovered by \emph{Einstein}, its large extent in
X-rays remained unnoticed prior to the \emph{ROSAT} observation.  The long
\emph{ROSAT} exposure and the low background made it possible to trace the
diffuse X-ray emission to a large radius and even perform a simple spectral
analysis. The optical and X-ray properties of 1159+5531 appear to be
representative of the other objects in our sample, and the main conclusions
derived from the 1159+5531 data can be applied to the entire OLEG
population.

The X-ray contour map of 1159+5531 overlayed on the optical CCD image is
shown in Fig~\ref{fig:1159:opt}. The X-ray surface brightness profile is
shown in Fig~\ref{fig:1159:x}a. The surface brightness is detected to
$1\,$Mpc from the center with $\approx 3\sigma$ significance. The
$\beta$-model fit (Cavaliere \& Fusco-Femiano 1976) in the entire radial
range yields $\beta=0.61$ and a small core-radius value $r_c=43\,$kpc.  The
central gas density derived from this fit corresponds to the gas cooling
time $t_{\rm cool}=2\times10^9\,$yr indicating the presence of the cooling
flow. Outside the central 200~kpc region, where the surface brightness is
unaffected by enhanced cooling flow emission, the best fit $\beta$-model
yields $r_c=400\pm185\,$kpc and $\beta=0.9\pm0.4$.  These large
uncertainties arise from the interplay between $r_c$ and $\beta$.
Fortunately, both gas and total mass at large radius depend mostly on the
slope of the surface brightness profile and not on the individual values of
$r_c$ and $\beta$. The surface brightness slope is well-constrained --- a
power law fit, $S\propto r^{-\gamma}$, between $600$ and $3000$\,kpc yields
$\gamma=3.6\pm0.6$.


Figure~\ref{fig:1159:x}b presents the gas temperatures in four annuli
measured by fitting the \emph{ROSAT} spectrum with the Raymond \& Smith
(1977) model. We fixed the metal abundance at $0.3$ of the Solar value and
the Galactic hydrogen column at $N_H=1.4\times10^{20}$\,cm$^{-2}$ derived
from the radio surveys (Stark et al.\ 1992). Fit to the overall \emph{ROSAT}
spectrum with free absorption yielded a consistent value, $N_H=1.5\pm0.4
\times 10^{20}$\,cm$^{-2}$. Within 100\,kpc, the temperature is
significantly lower than in the 100--250~kpc annulus, confirming the
existence of the cooling flow. Temperatures are poorly constrained beyond
250\,kpc because of low statistics and because \emph{ROSAT} is insensitive
above 2\,keV; nevertheless, the lower limits show that the temperature does
not drop below 1\,keV to at least 800\,kpc from the center. The temperature
$T=2.25\,$keV derived from the $\Lx-T$ correlation for 1--3\,keV clusters
and groups (Fukazawa 1997, Hwang et al.\ 1999) is consistent with the
spectral fit outside the cooling flow region. We will use this value for the
mass calculations below.

The gas mass was calulated by deprojection of the observed X-ray surface
brightness profile (Fabian et al.\ 1981, White, Jones \& Forman 1997). The
total mass was derived from the hydrostatic equilibrium equation assuming
constant temperature at all radii
\begin{equation}\label{eq:hydro-eq}
M(<R_{\rm x})=1.1\times10^{14} M_\odot\; T_{\rm keV} \; \beta
\;R_{\rm x}^3/\left(R_{\rm x}^2+r_c^2\right),
\end{equation}
where radii are in units of Mpc. The results of the mass determination at
several radii are presented in Table~\ref{tab:m:1159}. The formal
statistical uncertainty of the total mass is 30--40\%, including
uncertainties both in the slope of the surface brightness profile and in the
overall temperature.

\begin{inlinetable}
\caption{Total mass, gas mass, and optical luminosity of 1159+5531}
\label{tab:m:1159}
\begin{tabular}{p{1.25cm}ccccc}
\hline
\hline
\multicolumn{1}{c}{Radius}
& $M_{\rm tot}$ & $M_{\rm gas}$ & $L_{\rm opt}$ & $M_{\rm tot}/L$ & $f_{\rm gas}$\\
\multicolumn{1}{c}{(kpc)}& ($M_\odot$)  &  ($M_\odot$)  &  ($L_\odot$)  & 
   ($M_\odot/L_\odot)$\\
\hline
400\dotfill  & $4.6\times10^{13}$ & $3.9\times10^{12}$ & $4.2\times10^{11}$ &110&0.08\\
700\dotfill  & $1.2\times10^{14}$ & $1.0\times10^{13}$ & $4.4\times10^{11}$ &270&0.08\\
1000\dotfill & $2.0\times10^{14}$ & $1.4\times10^{13}$ & $4.6\times10^{11}$ &435&0.07\\
\hline
\end{tabular}
\end{inlinetable}

\begin{inlinefigure}
\includegraphics{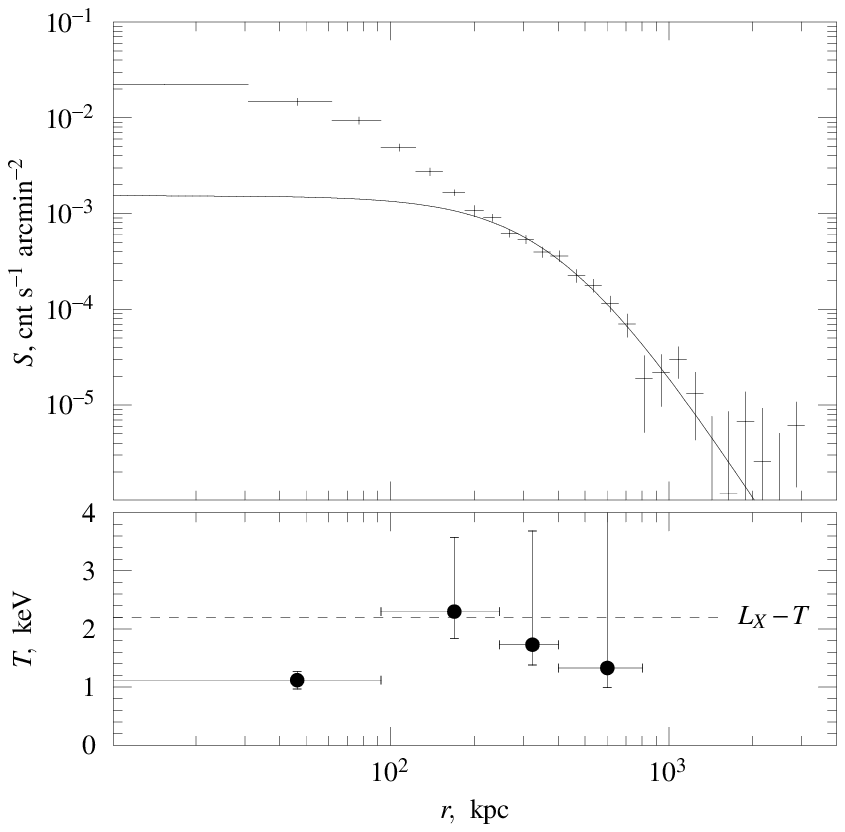}
\caption{{\em ROSAT}\/ PSPC
  X-ray surface brightness profile and temperature profile of 1159+5531.
  Solid line represents the $\beta$-model fit obtained excluding the central
  200~kpc affected by the cooling flow.}\label{fig:1159:x}
\end{inlinefigure}

Optically, 1159+5531 appears as an isolated giant elliptical residing at the
peak of the X-ray emission (Fig~\ref{fig:1159:opt}). The sensitivity limit
of this image corresponds to absolute magnitude $M_R=-14.8+5\lg h$ at the
object redshift. Above this sensitivity limit, we detect 18 galaxies within
a projected distance of 100~kpc from the central galaxy, while $15\pm4$
are expected due to the background estimated beyond 1~Mpc.

To measure the total optical luminosity, we integrated the light within
90~kpc of the central galaxy and within 15~kpc of fainter galaxies. We
excluded the objects with peak surface brightness exceeding that of the
central galaxy; the radial profiles of exluded objects showed that all they
were stars. The light density due to background galaxies was measured
outside a projected distance of 1~Mpc from the central galaxy.  The optical
luminosities within 0.4, 0.7, and 1~Mpc with the subtracted background
contribution are reported in Table~\ref{tab:m:1159}.

The optical properties of the central galaxy itself are remarkable. The
diffuse optical light can be traced to $\approx 150\,$kpc in $R$, $V$, and
$B$ bands. The light profile very accurately follows the de~Vaucouleurs law
with an effective radius $r_e=16\,$kpc. The absolute magnitude is
$M_R=-23.1 + 5 \lg h$ and $M_V=-22.5+5\lg h$ (with K- and Galactic
extinction correction applied). Such a high luminosity is rare in field
ellipticals but is typical of cD galaxies in clusters (Hoessel 1980).

To summarize, 1159+5531 is an object with X-ray properties typical of poor
clusters and which contains an isolated galaxy whose optical properties
resemble central cluster galaxies.

\section{Crude Mass Estimates in Other Objects}

A detailed analysis of the surface brightness and spectral fitting could be
performed only with the 1159+5531 data. In the remaining three objects, only
crude estimates of the total and gas mass are possible. To control the
accuracy of these estimates, we apply them also to 1159+5531. The gas
temperature is estimated from the $\Lx-T$ correlation (Fukazawa 1997, Hwang
et al.\ 1999). The temperature scatter around the mean $\Lx-T$ relation is
only $\approx25\%$, which is accurate enough for our purposes. Additional
confidence is added by the agreement of the $\Lx-T$ temperature estimate and
the direct spectral fit in $1159+5531$.  The X-ray surface brightness
profiles are fit with the $\beta$-model with free normalization and
core-radius, but we fix $\beta=0.67$ because of low statistics. Using the
$\beta$-model fit, we derive the central gas cooling time. In all objects
except for $2247+0337$, it is much shorter than the Hubble time, therefore
they likely contain cooling flows.  The gas mass within $R_{\rm x}$, the
radius of the detectable X-ray emission, is also derived using the
$\beta$-model fit. For a fixed $\Lx$, the derived value of gas mass is
relatively insensitive to $\beta$, therefore fixing $\beta=0.67$ does not
introduces significant errors in the gas mass.  Unfortunately, the presence
of the cooling flow limits the accuracy of the $\beta$-model gas mass
estimates. For example, a detailed modelling of the $1159+5531$ surface
brightness profile including the cooling flow yields $\approx 50\%$ larger
gas mass than the global $\beta$-model fit. The estimated temperature and
$\beta=0.67$ is substituted into eq~(\ref{eq:hydro-eq}) to derive the total
mass. The estimated gas and total masses are listed in
Table~\ref{tab:sample}. For $1159+5531$, the crude method yields a 30\%
lower gas mass and 20\% lower total mass compared to a more detailed
analysis in \S~\ref{sec:1159}.

\section{Discussion}

\subsection{Are OLEGs Really Isolated?}

Central galaxies in OLEGs clearly dominate their surroundings. The brightest
galaxy within the projected distance of 500\,kpc around $1159+5531$ is 2.9
magnitudes fainter and may be at a different redshift.  OLEGs do not show
any detectable concentration of galaxies in projection down up to 7.5
magnitudes fainter than the central galaxy. However, we cannot completely
exclude the existence of a dwarf galaxy population in these systems.
Mulchaey \& Zabludoff (1998) have studied NGC~1132, a nearby optically
selected galaxy with X-ray properties similar to OLEGs and $M_R=-21.5+5\lg
h$, and found a concentration of dwarf galaxies with $M_R$ in the range from
$-15$ to $-17+5\lg h$, consistent in number with that in X-ray detected
galaxy groups.  The sensitivity of our $1159+5531$ image is adequate to
detect such galaxies. We do not find any evidence for existence of a dwarf
concentration around $1159+5531$, but the upper limit on their number is
consistent with the composite group profile presented by Mulchaey \&
Zabludoff. The possibility of existence of a dwarf population around
$1159+5531$ and our other objects can be tested only by a detailed redshift
survey.

\subsection{Number Density}\label{sec:Spatial-Density}

The volume covered by our survey contains 4 objects. The corresponding
Bayesian lower and upper 95\% confidence limits of the true number of
objects are 1.97 and 9.15, respectively (see, e.g., Kraft, Burrows, \&
Nousek~1992). For given $\Lx$, the survey volume can be calculated using the
dependence of solid angle on limiting flux (Paper I). We find a volume of
$1.32\times10^{7}\,$Mpc$^{3}$ for objects with
$\Lx=2\times10^{43}\,$erg~s$^{-1}$; a slightly larger volume of
$1.67\times10^{7}\,$Mpc$^{3}$ is obtained for objects with
$\Lx=4\times10^{43}\,$erg~s$^{-1}$. Conservatively assuming the larger
volume, we obtain the spatial density of OLEGs of
$n=2.4^{+3.1}_{-1.2}\times10^{-7}\,$Mpc$^{-3}$ at 90\% confidence.

This number density is comparable to the number of other objects of similar
nature --- galaxy groups (compact groups in particular) and field elliptical
galaxies in the corresponding magnitude range. The X-ray luminosity
functions from Ebeling et al.\ (1997) and Burns et al.\ (1996) show that
OLEGs represent $\approx 20\%$ of all clusters and groups with
$\Lx>2\times10^{43}\,$erg~s$^{-1}$. Using the X-ray and optical luminosity
functions of Hickson compact groups (HCGs) from Ponman et al.\ (1996) and
Sulentic \& Raba\c ca (1994), we find that OLEGs outnumber comparably X-ray
luminous HCGs by a factor of $\approx 3.5$ and are as numerous as HCGs of
comparable total optical luminosity.  To estimate the number density of
field elliptical galaxies, we used the $R$-band luminosity function derived
from the Las Campanas redshift survey (Lin et al.\ 1996).  Above a limiting
absolute magnitude $M_R = -22.5+5 \lg h$ corresponding to the luminosity of
our faintest object $2114-6800$ (and 2.2\m\ brighter than $M^*$), we find
the number density of \emph{all} galaxies $n=2.50\times10^{-7}$\,Mpc$^{-3}$.
Most of these bright galaxies can be spectrally classified as early types
(Bromley et al.\ 1998). Some of them should lie within rich clusters, so the
above number density is an upper limit for the field population. Therefore,
most, if not all, of the brightest field ellipticals (with $L>6L^*$) are
OLEGs, i.e.\ possess X-ray halos extending for hundreds of kpc. This also
means that our OLEG sample cannot be significantly incomplete.

The contribution of OLEGs to the total mass in the Universe is close to that
of rich clusters. Markevitch (1998) derived the number density
$2.9\times10^{-8}\,$Mpc$^{-3}$ for $T>7\,$keV clusters. Since the total mass
scales with temperature as $T^{3/2}$, OLEGs ($T\approx 2\,$keV) are 6.5
times less massive than 7\,keV clusters. They outnumber such clusters by a
factor of $\approx 8$, and hence their contribution to the total mass
density is similar.

\subsection{High Mass-to-Light Ratio}

OLEGs have unusually high mass-to-light ratios. Within 500--1000~kpc, the
radius where extended X-ray emission can still be detected, $\ML$ in the
$R$-band is in the range 250--450$\,M_\odot/L_\odot$. We find similar or
slightly higher values of $\ML$ in the $V$-band. Our $\ML$ estimates are
2--3 times higher than the $V$-band values for clusters and groups
determined from the X-ray mass measurements (David, Jones \& Forman 1995).
Is this difference real or it can be explained by our measurement errors?
The largest uncertainty in $M/L$ is due to the inaccuracy of the total mass
measurement. Our total mass estimates are essentially proportional to
temperatures estimated from the $\Lx-T$ relation. A 2--3-fold overestimate
of $T$ from this relation is possible, but unlikely, given the good
agreement between the estimate and measurement for $1155+5531$.

A plausible explanation of the origin of OLEGs is that they are merged
compact galaxy groups (Ponman et al.\ 1994, Mulchaey \& Zabludoff 1998). If
this explanation is correct, our results imply that most of the brightest
field ellipticals are products of group merging and that these products are
more numerous than present-day groups. The latter is consistent with the
short estimated lifetime of compact groups (Mamon 1986).  However, the high
values of $M/L$ we find in OLEGs pose a problem for this scenario because
the optical luminosity is unlikely to decrease during the galaxy merging.

\section{Conclusions}

Even the scarce available data indicate that OLEGs represent a new
interesting class of objects warranting a detailed X-ray and optical study.
These objects are likely to have been undisturbed for a long time, and thus
represent the ultimate example of cluster-like systems in hydrostatic
equilibrium. Forthcoming \emph{Chandra} observations of two objects from our
sample will result in accurate profiles of the gas and total mass. If high
$M/L$ and low gas fraction is confirmed, this will pose a problem for low
$\Omega_0$ estimates from the values of these quantities in clusters.

\acknowledgements 

We thank M.~Markevitch for useful discussions. This work was supported by
the Smithsonian Institution and NAS8-39073 contract. HQ acknowledges support
from \mbox{FONDECYT} grant 8970009 and the award of a Presidential Chair in
Science.

\end{document}